\documentclass[aps,prd,twocolumn,floatfix,superscriptaddress,balancelastpage,nofootinbib,amsmath,showpacs]{revtex4}
\usepackage{graphicx}
\usepackage{epsfig}
\usepackage{bm}

\def\lsim{\mathrel{\raise.3ex\hbox{$<$\kern-.75em\lower1ex\hbox{$\sim$}}}}
\def\gsim{\mathrel{\raise.3ex\hbox{$>$\kern-.75em\lower1ex\hbox{$\sim$}}}}

\begin{document}

\hspace*{130mm}{\large \tt FERMILAB-PUB-12-596-A}

\title{On The Origin of IceCube's PeV Neutrinos} 

\author{Ilias Cholis}
\email{cholis@fnal.gov}
\affiliation{Center for Particle Astrophysics, Fermi National Accelerator
Laboratory, Batavia, IL 60510,USA}

\author{Dan Hooper}
\email{dhooper@fnal.gov}
\affiliation{Center for Particle Astrophysics, Fermi National Accelerator
Laboratory, Batavia, IL 60510,USA}
\affiliation{University of Chicago, Department of Astronomy and Astrophysics,
Chicago, IL, 60637, USA}

\begin{abstract}
\noindent 

The IceCube collaboration has recently reported the observation of two events with energies in excess of 1 PeV. While an atmospheric origin of these events cannot be ruled out at this time, this pair of showers may potentially represent the first observation of high-energy astrophysical neutrinos. In this paper, we argue that if these events are neutrino-induced, then the neutrinos are very likely to have been produced via photo-meson interactions taking place in the same class of astrophysical objects that are responsible for the acceleration of the $\sim$$10^{17}$ eV cosmic ray spectrum. Among the proposed sources of such cosmic rays, gamma-ray bursts stand out as particularly capable of generating PeV neutrinos at the level implied by IceCube's two events. In contrast, the radiation fields in typical active galactic nuclei models are likely dominated by lower energy (UV) photons, and thus feature higher energy thresholds for pion production, leading to neutrino spectra which peak at EeV rather than PeV energies (models with significant densities of x-ray emission, however, could evade this problem). Cosmogenic neutrinos generated from the propagation of ultra-high energy cosmic rays similarly peak at energies that are much higher than those of the events reported by IceCube.

\end{abstract}


\maketitle

\section{Introduction}

Very recently, the IceCube collaboration announced their observation of two events which could potentially represent the first detection of high-energy astrophysical neutrinos~\cite{two}. The analysis under consideration was designed to search for very high-energy ($\gsim$ PeV) and high-quality shower events over a period of 615.9 live days between 2010 and 2012.  Over this time, the IceCube experiment was in its nearly complete (79 string) and complete (86 string) configurations. The two showers in question were observed on August 8, 2011 and January 3, 2012, and each have an energy of approximately 1 PeV (1.04 and 1.14 PeV "with a combined and statistical uncertainty of $\pm 15 \%$ each" \cite{two}). Both showers were fully contained within the volume of the detector, and there are no indications of any instrumental problems, or of any connection with atmospheric muons. 

At energies below $\sim$1 PeV, atmospheric neutrinos constitute IceCube's primary background. Conventional atmospheric neutrinos, which are produced in the decays of pions and kaons present in cosmic ray induced cascades in the Earth's atmosphere, follow a spectrum which falls-off rapidly with energy (following a power-law spectrum with an index of approximately 3.7). Although this background is negligible at energies above several PeV, it cannot be entirely discounted in the case of the analysis at hand. In particular, the IceCube collaboration expects 0.082$\pm 0.004$(stat.)$^{+0.041} _{-0.057}$(syst.) background fully contained events with energy 1or more  PeV (including neutrinos from atmospheric muons, form decays of pions and kaons and prompt atmospheric neutrinos from decays of charmed mesons) over the time period covered by their analysis. The observation of two events with an expected background of 0.082 constitutes a P-value of $2.9 \times 10^{-3}$, corresponding to a significance of 2.8$\sigma$. 

In addition to the so-called conventional atmospheric neutrinos, a somewhat harder spectrum of prompt atmospheric neutrinos produced in the decays of heavy flavor (charm, bottom) mesons is also expected. Estimates of the event rate from this component of the atmospheric neutrino spectrum are less certain, although the lack of any correlation between the two observed events and hits in the IceTop surface array appears to disfavor this interpretation (efforts to quantify IceTop's veto efficiency are currently ongoing).

Given the quite modest statistical significance (2.8$\sigma$) represented by these two events, we cannot at this time be confident that they are of cosmic origin. That being said, if they are in fact cosmic neutrinos, then a few more years of data should reveal several more PeV showers. At a continued rate of 2 events per 615.9 live days, we estimate that this signal will become inconsistent with an atmospheric origin at the 5$\sigma$ level after a total of approximately 7 years of data taking. Searches for muon-track events or partially contained showers could also shed a great deal of light on the question of whether these events in fact originate from beyond the Earth's atmosphere. 

In this paper, we assume that this pair of intriguing events are astrophysical in nature and discuss the types of astrophysical sources from which they could potentially originate.  In Sec.~\ref{general}, we discuss in general terms the astrophysical production of PeV neutrinos and argue that these two events -- if, in fact, astrophysical in nature -- are likely to originate from the same class of objects that produce the bulk of the $10^{17}$ eV-scale cosmic rays. In particular, for such sources to produce the PeV neutrino flux implied by IceCube's two events, it is only necessary that on the order of ten percent of the energy in $\sim$$10^{17}$ eV cosmic rays is lost to photo-meson interactions. In this context, gamma-ray bursts are a particularly attractive possibility, which we discuss in Sec.~\ref{grb}. We consider other possible sources such as active galactic nuclei in Sec.~\ref{other}, but find these scenarios less compelling unless the objects in question are surrounded by significant densities of high energy (x-ray) radiation. We also present arguments for why these neutrinos are unlikely to be produced in the propagation of cosmic rays (cosmogenic neutrinos) or result from the decays of ultra-high energy neutrons. In Sec.~\ref{predictions}, we make predictions for future observations at IceCube and, in Sec.~\ref{summary}, we summarize our results and conclusions.

\section{The Astrophysical Production of PeV Neutrinos}
\label{general}

PeV-scale astrophysical neutrinos can be produced through three primary processes. Firstly, proton collisions with energetic photons can generate charged pions (photo-meson production), which yield neutrinos in their decays. To exceed the threshold for pion production, however, one must consider circumstances in which the target radiation is fairly energetic (10's of eV or above). Secondly, collisions between energetic protons (or neutrons) and gas can easily exceed the threshold for pion production. In sources of protons with enough energy to yield PeV neutrinos, however, the number density of sufficiently energetic photon targets present will almost certainly be much larger than that of nucleons. Thus we only expect $pp$ collisions to dominate neutrino production in sources of lower energy neutrinos \footnote{Above the pion production threshold $\sigma_{p\gamma} \simeq 0.2-0.4$ mb, while bellow the pion production threshold that cross-section is highly suppressed. The inelastic pp collisions cross-section is relatively constant at PeV energies being  $\sigma_{inel. pp} =$ 59 mb for a 1 PeV CR proton and 72 mb for a 10 PeV CR proton colliding with an ISM gas proton. Our arguments depend on the mean free path of a CR proton interacting with a target photon or proton, $l_{p\gamma}$, $l_{pp}$ with $l_{p\gamma} = 1/(\sigma_{p\gamma}\cdot n_{\gamma})$ and $l_{pp} = 1/(\sigma_{inel. pp}\cdot n_{p})$. $n_{\gamma}$, $n_{p}$ are the target photon and proton densities respectively.}. 
And lastly, PeV electron anti-neutrinos can result from the decays of ultra high-energy neutrons. As only a small fraction of a neutron's energy goes into its neutrino decay product, however, neutron decay anti-neutrinos are not predicted to be produced in sufficient numbers to account for the two events recently reported by the IceCube collaboration (see, for example, Refs.~\cite{Roulet:2012rv,Anchordoqui:2007fi}).

With these considerations in mind, we expect cosmic ray collisions with energetic photons to be the dominant mechanism behind IceCube's two reported events (assuming they are in fact astrophysical and not terrestrial in nature). And while one could also contemplate more exotic sources, such as the decays of ultra-heavy particles, we confine ourselves to conventional astrophysical scenarios in this paper.

Each of the three production mechanisms described above involve very high-energy cosmic rays, providing us with a direct connection between the observation of high-energy neutrinos and the observed cosmic ray spectrum~\cite{review}. More specifically, one can place an upper bound on the diffuse neutrino flux that results from cosmologically distributed, optically-thin cosmic ray accelerators. This argument, originally presented by Waxman and Bahcall in connection with cosmic rays of the highest energies~\cite{wb}, begins with the (cosmologically) local energy injection rate of ultra high-energy ($10^{19}-10^{21}$ eV) cosmic rays:
\begin{equation}
\left.E^2_{\rm{CR}} \frac{d\dot{N}_{\rm{CR}}}{dE_{\rm{CR}}}\right|_{10^{19}\,{\rm eV}} 
= \frac{\dot \epsilon_{\rm CR}^{[10^{19}, 10^{21}]}}{\ln(10^{21}/10^{19})} 
\approx 10^{44}\,\rm{erg}\,\rm{Mpc}^{-3} \rm{yr}^{-1}\,,
\end{equation}
where an injected energy spectrum $\propto$\,$E^{-2}$ has been assumed. The energy density of neutrinos produced through
the photo-meson interactions of these protons can be directly tied to the
injection rate of cosmic rays:
\begin{equation}
E^2_{\nu} \frac{dN_{\nu}}{dE_{\nu}}
\approx \frac{3}{8} \epsilon_\pi \, t_{\rm{H}}\,E^2_{\rm{CR}} 
\frac{d\dot{N}_{\rm{CR}}}{dE_{\rm{CR}}},
\end{equation}
where $t_{\rm{H}}$ is the Hubble time and $\epsilon_\pi$ is the
fraction of the energy which is injected in protons lost to photo-meson
interactions.  The factor of 3/8 comes from the fact that, near the threshold for pion production, roughly half the pions produced in photo-meson interactions are neutral and do not
generate neutrinos, and three quarters of the energy of charged pion
decays ($\pi^+ \rightarrow \mu^+ \nu_{\mu} \rightarrow e^+ \nu_e
\nu_{\mu} \bar{\nu}_{\mu}$) go into neutrinos. 

Taken together, the neutrino flux connected to the observed cosmic ray spectrum is given by:
\begin{eqnarray} [E^2_{\nu} \Phi_{\nu}]_{\rm WB} & \approx & (3/8)
  \,\xi_Z\, \epsilon_\pi\, t_{\rm{H}}\, \frac{c}{4\pi}\,E^2_{\rm{CR}}
  \frac{d\dot{N}_{\rm{CR}}}{dE_{\rm{CR}}} \nonumber \\ & \approx & 2.3
  \times 10^{-8}\,\epsilon_\pi\,\xi_Z\, \rm{GeV}\,
  \rm{cm}^{-2}\,\rm{s}^{-1}\,\rm{sr}^{-1},
\label{wbproton}
\end{eqnarray}
where the parameter $\xi_Z$ accounts for the effects of redshift dependent source evolution. 
Waxman and Bahcall originally presented this argument as an upper bound, derived for the case of $\epsilon_\pi=1$ (representing the maximum flux from a class of optically thin sources).\footnote{This result was derived specifically for photo-meson interactions and would be increased by a factor of $4/3$ if $pp$ interactions had instead been assumed.} 
At production, the neutrino flux from positively charged pion decays consists of equal fractions of $\nu_e$, $\nu_{\mu}$ and $\bar{\nu}_{\mu}$. After oscillations are taken in account, however, the muon neutrinos and anti-neutrinos become a roughly equal mixture of muon and tau flavors. 
For a distribution of sources which follows the star formation rate \cite{Yuksel:2008cu}, source evolution increases the normalization of the Waxman-Bahcall flux by a factor of $\xi_Z =$ 5.75 (see also \cite{wb}). We note that if we had used the GRB source evolution model as described in \cite{Yuksel:2006qb}, this factor would be larger, about 22.

It should be noted that if we drop the assumption that the cosmic ray injection spectrum is of the form, $dN_{\rm CR}/dE_{\rm CR} \propto E^{-2}_{\rm CR}$, we could potentially increase the flux of 1-10 PeV cosmic neutrinos, although not by more than a factor of about two~\cite{rpm}. More significantly, if the sources of the ultra high-energy cosmic rays are not optically thin to photo-meson production, as assumed in the derivation of the Waxman-Bahcall bound, then the observed cosmic ray spectrum could be made up of only the tail of the distribution of accelerated cosmic rays which escape from their sources. In such a case, much or even most of the energy that goes into accelerating ultra high-energy cosmic rays could be lost to the source environment, reabsorbing most would-be cosmic rays before they escape. From a class of such optically thick ``hidden'' sources, neutrino fluxes could plausibly be produced that are in excess of the Waxman-Bahcall bound by more than an order of magnitude, without exceeding the observed cosmic ray spectrum or gamma-ray background~\cite{rpm}.

So far, this calculation has assumed that the cosmic ray spectrum consists of only protons, rather than iron or other nuclei species. Heavy nuclei cosmic rays only produce charged pions after disintegrating into their constituent nucleons. Thus if the cosmic ray spectrum were dominated by heavy nuclei, the resulting neutrino flux would be further suppressed relative to the maximum value found in the standard Waxman-Bahcall calculation~\cite{Anchordoqui:2007tn}. In the cosmic ray  energy range around $\sim$$10^{17}$ eV that leads to the production of $\sim$1-10 PeV neutrinos, however, there is evidence that the cosmic ray spectrum is dominated by protons (although this appears to gradually evolve to a heavy nuclei dominated spectrum above $10^{19}$ eV)~\cite{Abraham:2009dsa,Hooper:2009fd}. We thus assume a proton dominated cosmic ray spectrum throughout this study.

For a high-energy cosmic neutrino spectrum, we can calculate the estimated rate of PeV-scale showers expected at IceCube. Hadronic showers can be generated through the neutral current interactions of all neutrino flavors  i.e $\nu_{l} N \rightarrow \nu_{l} X$, $\bar{\nu}_{l} N \rightarrow \bar{\nu}_{l} X$ ($X$ notating additional products)\footnote{Additionally in neutral current interactions there is the Glashow resonance \cite{Glashow:1960zz} $\bar{\nu}_{e} e \rightarrow W^{-} \rightarrow X$ at $\bar{\nu}_{e}$ energy of $E_{Gres} = M_{W}^2/(2m_{e})\simeq$6.3 PeV and a width of $\Gamma_{Gres} = E_{Gres} \Gamma_{W}/M_{W} \simeq 0.17 PeV$.}, with a typical shower energy that is about a quarter of that possessed by the initial neutrino. Alternatively, charged-current interactions of electron neutrinos and anti-neutrinos  ($\nu_{l} N \rightarrow l^{-} X$, $\bar{\nu}_{l} N \rightarrow l^{+} X$) produce a superposition of electromagnetic showers (evolving through the bremsstrahlung emission of a high energy photon from an electron/positron and through the subsequent $e^{+}e^{-}$ pair production per interaction length) and hadronic showers that in each interaction produce through the hadronization of quarks a wide variety of hadronic particles which subsequently decay into lighter particles including muons.  In the charged current interactions the electromagnetic and the hadronic showers collectively contain the entire energy of the incoming neutrino. While electromagnetic and hadronic showers are, in principle, distinguishable by their respective muon content, such a separation is generally expected to be difficult. In addition to producing tracks associated with charged leptons, charged current interactions of muon and tau neutrinos produce hadronic showers similar to those resulting from neutral current processes.

\begin{figure}[!]
%\vspace{-2.8cm}
\includegraphics[width=0.5\textwidth ]{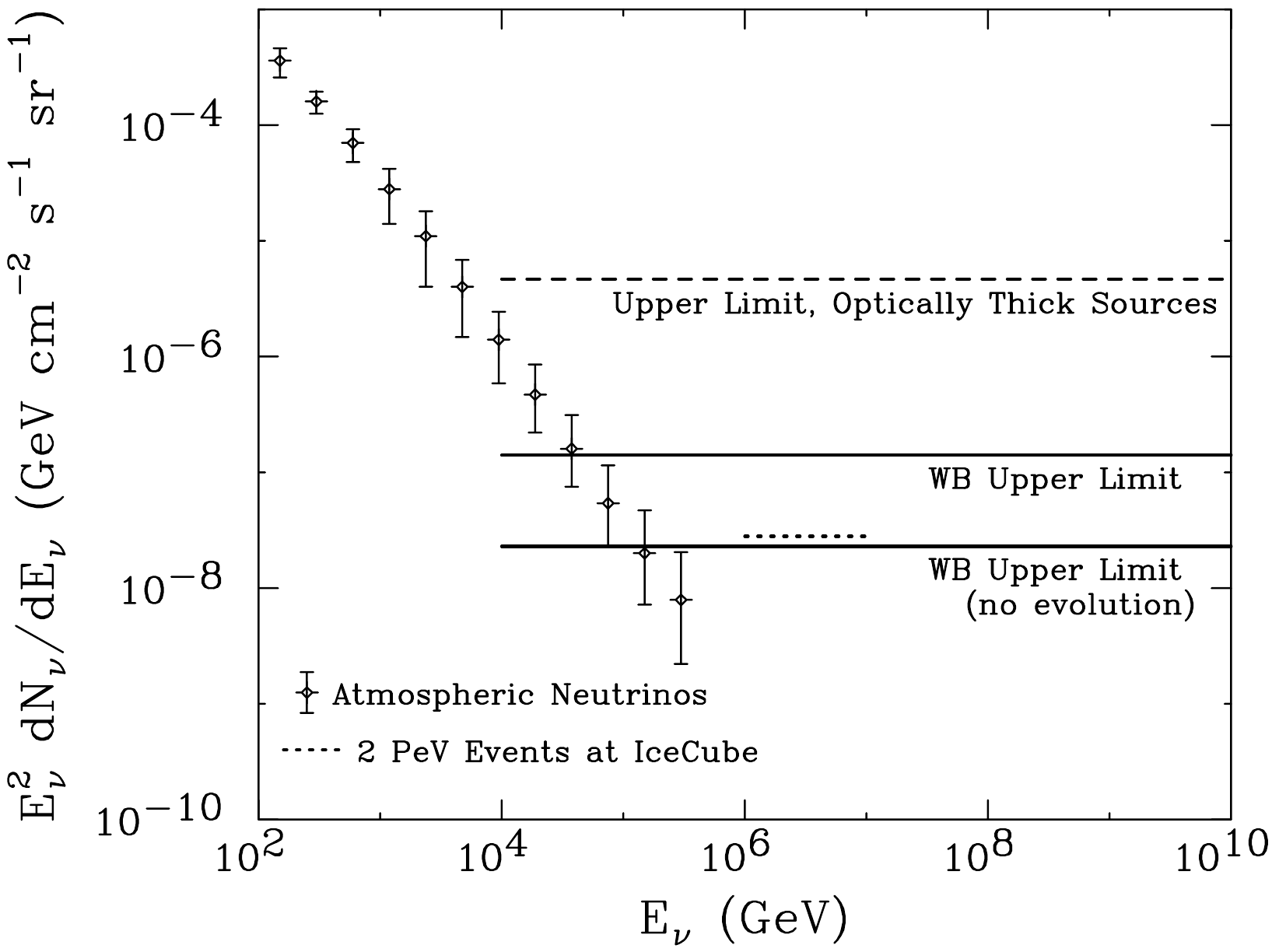}
%\vspace{-2.2cm}
\caption{To account for the two fully contained PeV shower reported by IceCube, a neutrino (plus anti-neutrino, all flavors) flux roughly at the level depicted by the dotted line is required. For comparison, we show (as solid lines) the Waxman-Bahcall upper limit on the diffuse neutrino flux~\cite{wb}, as derived for sources which follow a star formation rate-like redshift distribution, and for no redshift dependent source evolution. As a dashed line, we also show the less stringent limit derived for optically thick sources~\cite{rpm}. The error bars represent the spectrum of atmospheric neutrinos, as measured by IceCube~\cite{Abbasi:2010ie}.}
\label{wbfigure}
\end{figure}

PeV showers appear to the IceCube detector as photo-electrons distributed over an approximately $\sim$$300$ m radius sphere.  Although the two recently reported shower events were entirely contained within the volume of the experiment, IceCube should be capable of detecting partially contained showers as well. 

The probability that a neutrino passing through the effective area of IceCube produces an observable shower via a neutral current interaction is given by
\begin{equation}
P_{\nu \rightarrow \rm{shower}} 
\simeq \rho N_A L 
\int^{1}_{E_{\rm{sh}}^{\rm{thr}}/E_{\nu}} \frac{d \sigma}{d y} dy
\end{equation}
where $N_{A}$ the Avogadro number, $\sigma$ is the neutrino-nucleon cross section~\cite{cross}, $y$
is the energy fraction transferred from the initial neutrino to the target nucleus, $L$ is the length of the detector and $\rho$ is the density of ice.  For charged-current electron neutrino interactions, all of the neutrino energy goes into the shower, leading instead to $P_{\nu \rightarrow \rm{shower}} \simeq \rho N_A \sigma L$.

In calculating the expected rate of PeV-scale events at IceCube, we must also account for neutrino absorption in the Earth. A $\sim$1 PeV neutrino has only a few percent chance of passing through the equatorial diameter of the Earth without undergoing at least one interaction. In contrast, for neutrinos with an inclined trajectory of 10 degrees below the horizon, only about half of such particles will undergo one or more interactions in the Earth. The effects of absorption are negligible for downward-going neutrinos in the energy range under consideration. In calculating the effects of neutrino absorption, we adopt a simple density model of the Earth, consisting of a 2500 km radius core with a uniform density of 11,000 kg/m$^3$ and a uniform density outer region normalized to the overall mass and radius of the Earth.

For a high-energy cosmic neutrino spectrum that saturates the Waxman-Bahcall bound ($\epsilon_{\pi}=1$), from sources distributed without significant redshift evolution ($\xi_Z=1$), we calculate that IceCube should observe a rate of 13.6 showers per year, per cubic kilometer, with more than 1 PeV of energy. If we further require that the showers be initiated in the inner volume, restricting ourselves to fully contained events, the effective volume of IceCube is reduced to roughly $\sim$0.1 km$^3$, and thus we predict IceCube to observe 1 or 2 fully contained PeV-showers per year. 

In Fig.~\ref{wbfigure}, we compare the approximate neutrino flux required to explain the two PeV showers reported by the IceCube collaboration to the Waxman and Bahcall bound, both without source evolution and with source evolution that follows the star formation rate. We also show the less stringent upper bound derived in the case of optically thick sources~\cite{rpm}. The fact that the neutrino flux required to explain IceCube's two PeV showers lies not far below the Waxman-Bahcall bound implies that a significant fraction ($\gsim$10\%) of energy injected into $\sim$$10^{17}$ eV cosmic rays must be transferred into pions. We also note that IceCube's current (90\% CL) upper limit on the diffuse neutrino flux in this energy range is $2.7\times 10^{-8}$ GeV cm$^{-2}$ s$^{-1}$ sr$^{-1}$, assuming an $E^{-2}$ spectrum between $3.5 \times 10^4$ and $7\times 10^{6}$ GeV~\cite{Abbasi:2011jx}, which is very close to the flux required to account for the two reported events. 
As the majority of such events are expected to occur at energies below a few PeV, the precise spectral shape at higher energies impacts this rate only marginally.

In the following section, we move away from these general arguments, and consider gamma-ray bursts as a specific class of sources that is particularly well suited to produce the required flux of neutrinos in the  PeV energy range.

\section{PeV Neutrinos From Gamma-Ray Bursts}
\label{grb}

Gamma-ray busts (GRBs) constitute one of the most promising sources of high and ultra-high energy cosmic rays, and may be capable of accelerating protons to energies as high as $\sim$$10^{20}$ eV~\cite{Waxman:1995vg}. Furthermore, as their name implies, gamma-ray burst fireballs contain high densities of gamma-rays, enabling for the efficient production of neutrinos via the photo-meson interactions of high energy protons~\cite{waxman,meszaros}.

More specifically, typical GRBs exhibit a broken power-law spectrum of the form: $dN_{\gamma}/dE_{\gamma} \propto E_{\gamma}^{-2}$ for $E_{\gamma}\gsim 0.1-1$ MeV and $dN_{\gamma}/dE_{\gamma} \propto E_{\gamma}^{-1}$ at lower energies~\cite{band}. Furthermore, the radiation pressure resulting from the very high optical depth of GRB fireballs leads to their ultra-relativistic expansion, accelerating the plasma to Lorentz factors on the order of $\Gamma \sim 10^2-10^3$. In order for proton-photon collisions in this environment to exceed the threshold for pion production, the proton must have an energy (in the observer's frame) that meets the following condition~\cite{grb,guetta}:
\begin{equation}
E_p \gsim 40 \,{\rm PeV} \,\bigg(\frac{\Gamma}{300}\bigg)^2 \, \bigg(\frac{0.3 \, {\rm MeV}}{E_{\gamma}}\bigg)\, \bigg(\frac{1}{1+z}\bigg)^2,
\label{grbc1}
\end{equation}
where $z$ is the redshift of the burst. For any falling spectrum of high-energy protons, such interactions will predominantly take place near this threshold. After taking into account that only about 20\% of the proton's energy goes into the charged pion produced in such an interaction, and that each neutrino carries away only about a quarter of the charged pion's energy, this leads to the production of neutrinos of characteristic energy:
\begin{equation}
E_{\nu} \sim 2 \,{\rm PeV} \,\bigg(\frac{\Gamma}{300}\bigg)^2 \, \bigg(\frac{0.3 \, {\rm MeV}}{E_{\gamma}}\bigg)\, \bigg(\frac{1}{1+z}\bigg)^2.
\label{grbc2}
\end{equation}
Thus for protons interacting with photons near the observed spectral break, the resulting neutrinos will have energies near that of the two events reported by IceCube. The neutrino flux at energies below this value will be suppressed by the lack of sufficiently high energy target photons in the fireball. For this reason, the PeV energy scale is where one roughly expects to observe the first GRB neutrinos (see also \cite{Baerwald:2012kc}).

The overall normalization of the diffuse neutrino flux from all GRBs depends on how much of the bursts' internal energy goes into accelerating protons to energies of $\sim$$10^{16}$ eV and above. It has been appreciated for some time that if the majority of the highest energy cosmic rays originate from GRBs, then one should also expect an observable diffuse flux of high energy neutrinos~\cite{grb,guetta}. 
In our calculations, we adopt a ratio of ten between the energy that goes into accelerated protons and electrons~\cite{Zhang:2006uj, Racusin:2011jf}, and also assume that $\simeq$1\% of the energy in accelerated protons goes into neutrinos. There are significant deviations in that factor between individual GRBs~\cite{guetta}, depending on the fireball environment. A value of $10\%$ of the proton energy going to neutrinos corresponds to $\epsilon_{\pi} \approx 0.2$. Higher fractions of proton energy going to neutrinos would result in higher neutrino fluxes keeping all the other GRB assumptions fixed. 

Most observed GRBs exhibit maximum isotropic luminosities in the range of $L_{\rm max} \sim 10^{51}-10^{53}$ erg/s.\footnote{As the observed hard x-ray and $\gamma$-ray luminosity is synchrotron emission from internal shocks in the relativistic fireball~\cite{Rees:1994nw}, this emission will be relativistically beamed to within an opening angle on the order of $\theta \sim 1/\Gamma$. In our calculations, we use the isotropic equivalent luminosity related to the true luminosity by: $L_{\rm iso} = L_{\rm true}/(1-\cos\theta)$. This avoids overestimating the neutrino or photon fluxes. We also estimate the total isotropic energy emitted to be $E_{\rm iso} \simeq L_{\rm iso}^{\rm max} \tau_{\rm dur}$. $L_{\rm iso}^{\rm max} \equiv L_{\rm max}$(for simplicity in the remaining text) is the maximum isotropic equivalent luminosity;  The duration timescale as observed in hard x-rays and $\gamma$-rays, is taken to be  $\tau_{\rm dur} = 2$ sec for high luminosity GRBs and and $\tau_{\rm dur} = 50$ sec for the low luminosity sample. Here, $\tau_{\rm dur}$ is taken as the timescale between which 25\% and 75\% of the total energy (in x-rays and $\gamma$-rays) has been emitted (see \cite{Goldstein:2012uf, Paciesas:2012vs}).} Such bursts are referred to in literature as high luminosity GRBs and are further divided into short and long duration bursts with observed timescales of 0.1-1 and 10-100 seconds, respectively; with the majority of observed bursts being of long duration~\cite{Zhang:2012jr, Goldstein:2012uf, Paciesas:2012vs}. In addition, a few low luminosity GRBs with $L_{\rm max} \sim 10^{47}$ erg/s have been detected, potentially representing another distinct GRB population~\cite{Liang:2006ci} (see also Ref.~\cite{Lv:2010bz}). These low luminosity GRBs, which are potentially much more numerous than their high luminosity counterparts, generally exhibit smooth light curves, wider emission cones and longer durations (50-1000 s)~\cite{Murase:2006mm, Gupta:2006jm, Coward2006, Cobb:2006cu, Pian:2006pr, Daigne:2007qz, Bromberg:2011fm}. 
%%%

%%%

Both high and low luminosity GRB populations can be described by a luminosity distribution parametrized as:
\begin{equation}
\Phi(L) = \Phi_{0}\left[\left(\frac{L}{L_{b}}\right)^{\alpha_{1}} + 
\left(\frac{L}{L_{b}}\right)^{\alpha_{2}} \right]^{-1},
\label{eq:LuminDistr}
\end{equation}
where $\Phi_{0}$ normalizes the luminosity distribution to 1 (within 2 orders of magnitude above/below the luminosity break $L_{b}$). At present, there is considerable variation in the values of these parameters as they appear in the literature~\cite{Gupta:2006jm,Liang:2006ci,Wanderman:2009es}. In our calculations, we adopt ranges based on 2 years of \textit{Swift} data as presented in Ref.~\cite{Liang:2006ci}: $L_{b} = (1.2 \pm 0.6)\times 10^{52}$ erg/s, 
$\alpha_{1} = 0.65 \pm 0.15$, $\alpha_{2} = 2.3 \pm 0.3$ for high luminosity GRBs and 
$L_{b} = (1.0 \pm 0.3)\times 10^{47}$ erg/s, $\alpha_{1} = 0 \pm 0.5$, 
$\alpha_{2} = 3.5 \pm 0.5$ for low luminosity GRBs. These parameters lead to mean values (of the GRB distribution) for the (individual) GRB maximum isotropic luminosity of $\bar{L}_{max} = 3.7 \times 10^{51}$ to $3.8 \times 10^{52}$ erg/s for high luminosity GRBs and $\bar{L}_{max} = 3.5 \times 10^{46}$ to $1.6 \times 10^{47}$ erg/s for low luminosity GRBs. Throughout the remainder of this study, we will use as reference values $\bar{L}_{\rm max}^{HL} = 1 \times 10^{51}$ erg/s and $\bar{L}_{\rm max}^{LL} = 10^{47}$ erg/s, but will also consider variations in order to test the assumptions of Ref.~\cite{Gupta:2006jm} and~\cite{Liang:2006ci} (see also~\cite{He:2012tq} for an alternative analysis on the impact of different luminosity distribution assumptions). 

%%%

%%%

For the redshift distribution (co-moving rate density) of GRBs, we adopt the star formation rate of Ref.~\cite{Porciani:2000ag} (model ``SF2''):
\begin{equation}
R_{\rm GRB}(z) = 23 R_{\rm GRB}(0) \, \frac{e^{3.4 z}}{e^{3.4 z} + 22}.
\label{eq:RedshiftGRB1}
\end{equation}
High luminosity GRBs occur at a local rate on the order of 1 Gpc$^{-3}$yr$^{-1}$
\cite{Liang:2006ci, Coward:2012gn}, while the rate of low luminosity GRBs is considerably higher, with estimate ranging from 230~\cite{Soderberg:2006vh} to $5\times 10^{3}$~\cite{Soderberg:2005vp} Gpc$^{-3}$yr$^{-1}$.
% with the larger value constraint coming from SNe Ibc observations.
We take as reference values $R_{\rm GRB}^{HL}(0) = 1.0$ and $R_{\rm GRB}^{LL}(0) = 350$ Gpc$^{-3}$yr$^{-1}$.
While it has been suggested~\cite{Wanderman:2009es} that the GRB rate redshift distribution does not follow the star formation rate, but is instead suppressed at $z>3$, this would not affect the observed fluxes from high luminosity GRBs by more than about a factor of 2 from our reference values for high luminosity GRBs.

The diffuse flux of neutrinos or photons at the location of the Earth from the population of all GRBs is given by: 
\begin{eqnarray}
\frac{dN^{\rm obs}}{dE_{\nu,\bar{\nu}, ph}^{\rm obs}} &=& \int_{0}^{z_{\rm max}} dz \int_{L_{\rm min}}^{L_{\rm max}} dL  \;
\Phi(L) \frac{R_{\rm GRB}(z)}{1+z} \frac{4\pi D_{L}(z)^{2}}{(1+z)^2} \nonumber \\ 
&\times& \frac{c}{H_{0} \sqrt{\Omega_{\Lambda} + \Omega_{M}(1+z)^3}} \frac{dN^{\rm obs}}{dE^{\rm obs}},
\label{eq:Nuflux}
\end{eqnarray}
where 
\begin{equation}
D_{L}(z)/(1+z) = D(z) = \int_{0}^{z} \frac{c}{H_{0}}\frac{dz'}{\sqrt{\Omega_{\Lambda} + \Omega_{M}(1+z')^{3}}},
\label{eq:Dist}
\end{equation}
and $dN^{\rm obs}/dE^{\rm obs}$ refers to the observable neutrino/photon fluence from
an individual GRB located at comoving distance $D(z)$ (luminosity distance $D_{L}$):
\begin{equation}
\frac{dN^{\rm obs}}{dE^{\rm obs}} = \frac{dN^{\rm inj}}{dE^{\rm inj}} \frac{1+z}{4\pi D(z)^{2}}. 
\label{eq:NuObsSpect}
\end{equation}
$dN^{\rm inj}/dE^{\rm inj}$ is the equivalent injection neutrino/photon spectrum. We take as $z_{\rm max} =9.4$ 
\footnote{The exact value of $z_{\rm max}$ with $z_{\rm max} >7$ does not impact our calculations by more than $1\%$.}, which is the most distant GRB that has been detected~\cite{Cucchiara:2011pj}.

%Finally we account for the contribution of HL and LL GRBs separately. 

The spectrum of neutrinos (and anti-neutrinos) at injection can be approximated by a doubly broken power-law~\cite{Waxman:1998yy}:
\begin{equation}
\frac{dN_{\nu}^{\rm inj}}{dE_{\nu}^{\rm inj}} \propto \left\{ \begin{array}{ll} 
& \left(\frac{E_{\nu}}{E_{1}}\right)^{-1} \; \textrm{for} \; E_{\nu} \leq E_{1}   \\
&\left(\frac{E_{\nu}}{E_{1}}\right)^{-2} \; \textrm{for} \; E_{1} \leq E_{\nu} \leq E_{2}  \\
&\left(\frac{E_{2}}{E_{1}}\right)^{-2} \times \left(\frac{E_{\nu}}{E_{2}}\right)^{-3} \; \textrm{for} \; E_{\nu} \geq E_{2} 
\end{array}\right\}.
\label{eq:NuInjSpect}
\end{equation}
The first of these spectral features (at $E_{\nu}=E_1$) corresponds to the pion production threshold for scattering off photons at the observed break in the gamma-ray spectrum of GRBs, while the higher energy break (at $E_{\nu}=E_2$) appears as a result of the synchrotron cooling of muons and pions. The exact locations of these breaks is different for typical high and low luminosity GRBs due to differences in the strengths of the fireballs' magnetic and radiation fields. In calculating the location of the first break, we assume a gamma-ray spectrum which breaks at 1 MeV or 0.1 MeV for high and low luminosity GRBs, respectively. 
Furthermore the exact power-law index bellow our first break (located between $10^5$ and $10^6$ GeV) in the injected neutrino spectrum of eq.~\ref{eq:NuInjSpect} can not influence the flux above at 1 PeV, especially since the GRB neutrino fluxes are normalized by their injected energy to neutrinos dominated by the contribution of neutrinos with energy between the two breaks of eq.~\ref{eq:NuInjSpect}.
Deviations from the standard Waxman-Bahcall spectrum reproducible by the $\Delta^{+}$ resonance can result from additional neutrino production modes~\cite{Baerwald:2010fk}. These modifications of the spectrum once normalizing the injected energy to neutrinos -which is the main uncertainty- do not affect our calculations by more than a factor of 2.

\begin{figure}[!]
%\vspace{-2.8cm}
\includegraphics[width=0.5\textwidth ]{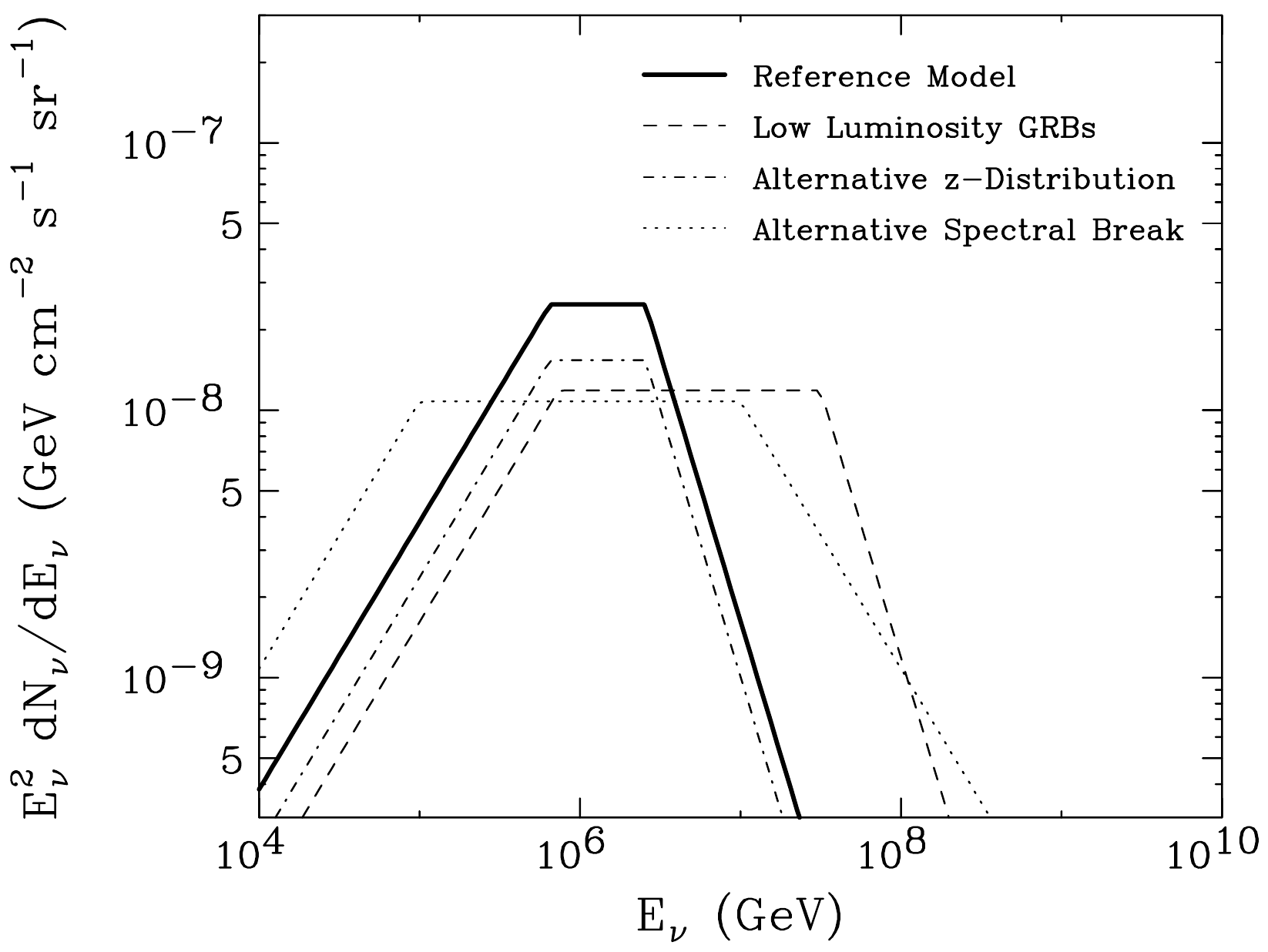}
%\vspace{-2.2cm}
\caption{The contribution of gamma-ray bursts (GRBs) to the diffuse neutrino (plus anti-neutrino) spectrum. Results are shown for high luminosity (solid) and low luminosity (dashed) GRBs, calculated using our default parameters, and for high luminosity GRB models with a suppressed high redshift distribution (dot-dash) and alternative spectral characteristics (dots). Each of these models yields a rate of PeV events which is comparable to that implied by the two events reported by IceCube. See text for details.}
\label{fig:GRBvariations}
\end{figure}

In Fig.~\ref{fig:GRBvariations}, we show the diffuse flux of neutrinos and anti-neutrinos from GRBs for our default parameter choices, and for some representative variations of these parameters. The solid and dashed lines represent the predicted flux for our default assumptions, from high and low luminosity bursts, respectively. The dot-dashed line represents the contribution from high luminosity GRBs with a redshift distribution which is suppressed above $z=3$. The dotted line shows the flux from high luminosity GRB with alternative choices for the parameters leading to the location of the spectral breaks. While these variations are by no means exhaustive, they demonstrate that for a fairly wide range of assumptions, GRBs are expected to generate fluxes of PeV neutrinos that are similar to that implied by IceCube's two events (see the dotted line in Fig.~1). In particular, for each of the four neutrino fluxes shown in Fig.~\ref{fig:GRBvariations}, one expects a rate of 4-7 showers with energies above 1 PeV per cubic kilometer, per year. For an estimate of $\sim$0.1 km$^3$ for the effective volume for fully contained showers, this range of rates is within about a factor of two of that required to account for the two events reported by IceCube.

Among the various uncertainties that are involved in the calculation of the neutrino flux from GRBs, a few stand out as particularly important. First of these is the fraction of the energy that goes into accelerating protons to energies above $10^{16}$ eV. In the case of the contribution from high luminosity GRBs, uncertainties in the luminosity function are particularly significant. For low luminosity GRBs, the local rate of such objects constitutes a major uncertainty. At present, these uncertainties can collectively impact expectations for the diffuse neutrino flux by as much as an order of magnitude and will continue to do so until they are better observationally constrained.

%%%%%

By restricting an analysis to events which correlate in time and/or direction to known GRBs, it is possible to conduct a nearly background free search for neutrinos originating from GRBs. Recently, the IceCube collaboration has applied such a strategy, and used the results to derive a stringent upper limit on the flux of high energy neutrinos from observed GRBs~\cite{Abbasi:2012zw}. Under standard astrophysical assumptions, this limit implies that GRBs cannot be the only sources of the highest energy ($>10^{18}$ eV) cosmic rays (see also, Ref.~\cite{Ahlers:2011jj}). The two events being considered in this paper, however, could still originate from GRB if either,  1) a greater fraction than expected of the high energy neutrinos from GRBs originate from bursts which are not sufficiently luminous to be observed by gamma-ray or x-ray observatories, or 2) a significant fraction of the $10^{16}-10^{18}$ eV cosmic ray spectrum (but not most of the $>10^{18}$ eV cosmic ray spectrum) originates from GRBs. This later possibility is attractive from the perspective of the cosmic ray spectrum's chemical composition. Measurements from the Pierre Auger Observatory of the depth of shower maxima and its variation suggest that the highest energy cosmic rays are largely of heavy chemical composition (closer in mass to iron nuclei than protons), while the composition becomes steadily lighter at lower energies, appearing to be dominated by protons at $10^{18}$ eV~\cite{heavy}. As ultra high-energy nuclei accelerated in a GRB are expected to be entirely disintegrated into individual nucleons before escaping the fireball~\cite{Anchordoqui:2007tn}, the possibility that GRBs provide much of the cosmic ray spectrum below $\sim$$10^{18}$ eV, but that another class of sources provide the bulk of the highest energy (heavy nuclei) cosmic rays, is a well motivated one.

The fact that the two PeV shower events reported by IceCube do not correlate in time with any known GRB does not significantly disfavor the hypothesis that these events originate from GRBs. The ability of gamma-rays and x-ray observatories to monitor for GRBs is, at present, substantially incomplete. In particular, the \textit{Swift} Burst Alert Telescope (BAT) and the \textit{Fermi} Gamma-ray Burst Monitor (GBM) collectively cover less than $\simeq$2/3 of the sky at any given time. With this in mind, one cannot rule out the non-negligible possibility that the two PeV shower events originated from GRBs which happened to fall outside of the combined field-of-view of these instruments. Furthermore, many GRBs, while in the field-of-view of either \textit{Swift}'s BAT or \textit{Fermi}'s GBM, may still go undetected if they are of sufficiently low luminosity, or are sufficiently distant.

To assess whether this later possibility is consistent with a GRB interpretation of these events, we must attempt to estimate the efficiency with which neutrino-producing GRBs are detected in the gamma-ray and x-ray bands~\cite{Paciesas:2012vs, Goldstein:2012uf}. Given the fluence sensitivity of the the \textit{Swift}'s BAT and \textit{Fermi}'s GBM $\sim$$1\times 10^{-8}$ erg cm$^{-2}$\,s$^{-1}$ and $\sim$$2\times 10^{-8}$ erg cm$^{-2}$\,s$^{-1}$ (at 20 keV) respectively, one can estimate how distant a GRB (of a given luminosity) could be and still trigger these detectors. We find that these experiments should be capable of detecting essentially all high luminosity GRBs ($L \gsim 10^{51}$ erg/s) within their fields-of-view out to a distance of about 8 Gpc $(z\approx 5)$. Thus the observed collection of high luminosity GRBs is fairly complete (within the given fields-of-view). In contrast, low luminosity GRBs ($L\sim 10^{47}$ erg/s) are likely to be detected only within a radius of $\sim$$100$ Mpc, suggesting that the vast majority of the diffuse neutrino flux from low luminosity GRB will not be correlated in time or direction with any observed gamma-ray or x-ray signal.

%%%%%%%%%%%%%%%%%%

\section{PeV Neutrinos From Other Astrophysical Sources}
\label{other}

\subsection{Active Galactic Nuclei}

\begin{figure}[!]
%\vspace{-2.8cm}
\includegraphics[width=0.5\textwidth ]{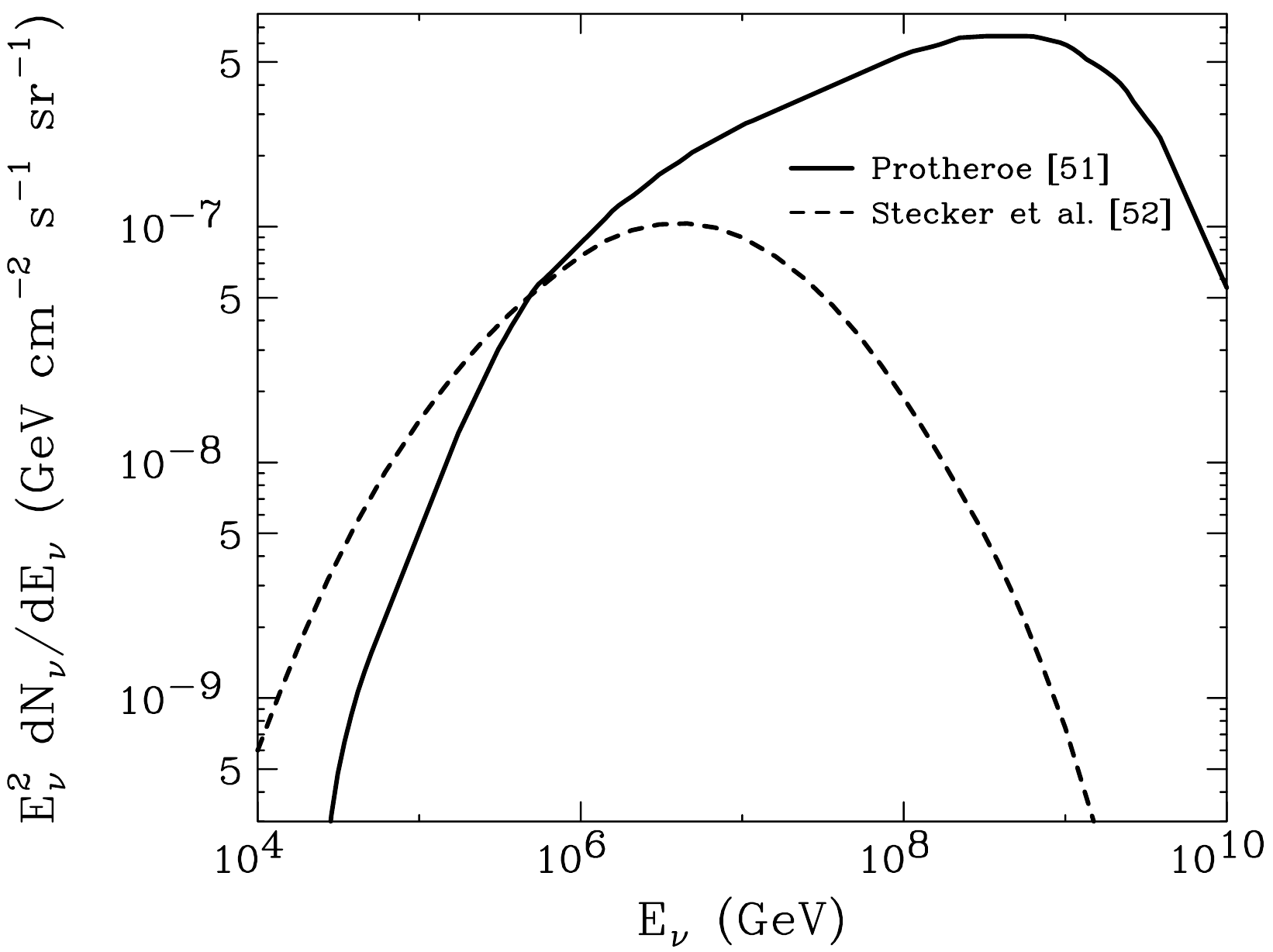}
%\vspace{-2.2cm}
\caption{The contribution of active galactic nuclei (AGN) to the diffuse neutrino (plus anti-neutrino) spectrum. Results are shown for the models of Protheroe~\cite{Mannheim:1998wp} and Stecker et al.~\cite{Stecker:1991vm}.}
\label{fig:AGN}
\end{figure}

The kinematics of high-energy neutrino production in Active Galactic Nuclei (AGN) is similar to that described for GRBs in Eqns.~\ref{grbc1}-\ref{grbc2}, but with potentially important differences. In particular, the Lorentz factors of AGN jets are significantly lower than those of GRB shocks; values of $\Gamma\sim 30$ rather than $\sim$300 are thought to be typical~\cite{agn}. As a result, $\sim$10-100 PeV protons can exceed the threshold for pion production much more easily, requiring only the presence of $\sim$keV photons (rather than the $\sim$100 keV photons required in GRBs). 

In GRBs, the observed photon spectral break ($\sim$0.1-1 MeV) leads to a break at $\sim$1 PeV in the neutrino spectrum. Thus we may expect the first detections of GRB neutrinos to appear at around this energy scale. In contrast, AGN do not typically exhibit a spectral peak at keV energies, but instead in the UV, typically at around $\sim$10 eV. This leads one to expect the neutrino spectrum to peak EeV energies, much higher than that from GRBs. There is a considerable degree of model dependence in this conclusion, however, deriving in large part from uncertainties in the spectrum of the target radiation fields.

We illustrate the nature of this uncertainty in Fig.~\ref{fig:AGN}, with a comparison of two canonical models of the diffuse neutrino emission from AGN. First, as a solid line, we show the diffuse neutrino spectrum as predicted by Protheroe~\cite{Mannheim:1998wp}. In this model, the scattering of ultra-high energy protons with UV radiation leads to a neutrino spectrum which peaks at EeV energies. For this spectral shape, most showers initiated within IceCube's volume will be of energy 20 PeV or greater. If IceCube's existing data does not contain at least a few enormous (non-contained) showers of this energy, this AGN model will not be able to account for the two reported PeV events. In contrast, the model of Stecker et al.~\cite{Stecker:1991vm} predicts a neutrino spectrum from AGN which peaks at a much lower energy of a few PeV, not unlike the predictions from GRBs. This is in large part due to the high density of ambient x-rays present in the Stecker {\it et al}. AGN model. 

We also point out that while neutrino emission from known GRBs can be efficiently constrained by searching in the time window around the occurrence of a given burst, such a background-free strategy is not possible for AGN. As a result, while it may be possible to rule out (high luminosity) GRB as the source of IceCube's two PeV events by searching in the time and direction of known GRBs, it will be much more difficult to definitively test the hypothesis that these neutrinos originate from AGN. 

\subsection{Starburst Galaxies}

Collisions of cosmic rays with the radiation in galaxies undergoing periods of rapid star formation are predicted to yield significant fluxes of $\sim$TeV-PeV neutrinos~\cite{starbursts}. For reasonable assumptions, and calibrating to the observed fluxes of gamma-rays~\cite{Lacki:2010vs} and radio emission from such objects, contributions to the diffuse neutrino flux are expected to be on the order of $\sim$$1\times10^{-8}$ GeV cm$^{-2}$ s$^{-1}$ sr$^{-1}$~\cite{Lacki:2010vs,starbursts,Stecker:2006vz}. While it is not clear whether this diffuse spectrum extends to energies as high as $\sim$PeV, if it does, starburst galaxies could potentially contribute at a level capable of accounting for the two PeV events reported by IceCube.

\subsection{Cosmogenic Neutrinos}

In order to generate $\sim$1-10 PeV neutrinos via the production and decay of charged pions, protons with energies on the order of of $\sim$$10^2$ PeV are required. Furthermore, in order to exceed the threshold for pion production in the center-of-mass frame, these ultra-energetic protons must encounter a sufficient density of energetic photons (or nucleons). A $10^{17}$ eV proton, for example, can only produce pions by scattering off of photons in excess of approximately 1 eV. The intergalactic background densities of radiation at such energies are insufficient to lead to any significant amount of pion production~\cite{Roulet:2012rv}, we restate the well known conclusion that the universe is transparent to $10^{17}$ eV protons, and thus find that cosmogenic neutrinos are unlikely to be the source of the two events reported by IceCube (see also \cite{DeMarco:2005kt, Ahlers:2010fw}).

We also note that while electron anti-neutrinos from the decays of ultra-high energy neutrons could, in principle, contribute to the diffuse neutrino flux at PeV energies, such neutrinos are insufficient to account for the events reported by IceCube, even in heavy nuclei dominated cosmic ray scenarios~\cite{Roulet:2012rv,Anchordoqui:2007fi}.

Finally, large scale intergalactic shock waves produced due to structure formation may be capable of accelerating both electrons and protons to Lorentz factors as high as $\sim$$10^{7}$~\cite{Keshet:2002sw, Gabici:2002fg}. Through inverse Compton scattering, the resulting cosmic ray electrons can produce up to $\sim$10\% of the observed isotropic gamma-ray spectrum above 50 GeV. The corresponding protons, are constrained to not produce through inelastic pp collisions significant fluxes of gamma-rays or neutrinos, unless the production of GeV scale $\gamma$-rays from CR electrons -setting the constraint for this sources- is highly suppressed (see~\cite{Murase:2008yt}).

\section{Predictions for Future Observations}
\label{predictions}

If the two PeV shower events observed by IceCube are, in fact, neutrino induced events, then one should expect them to be accompanied by a significant rate of corresponding muon track events. Once produced through the charged current interaction of a muon neutrino (or antineutrino), a muon loses energy at a rate given by:
\begin{equation}
\frac{dE_{\mu}}{dX}=-\alpha-\beta E_{\mu},
\end{equation}
where, in ice, $\alpha \approx 0.002 \, {\rm GeV} \,{\rm cm}^2/{\rm g}$ and $\beta \approx 4.2\times 10^{-6} \, {\rm cm}^2/{\rm g}$. Thus, before the energy of a propagating muon falls below $E^{\rm thr}_{\mu}$, it travels a distance of:
\begin{equation}
R_{\mu} \approx \frac{1}{\beta \rho}\log\bigg[\frac{\alpha+\beta E_{\mu}}{\alpha+\beta E^{\rm thr}_{\mu}}\bigg].
\end{equation}
For PeV-scale muons, this range will significantly exceed the linear scale of IceCube, increasing the effective target mass of the experiment. The probability of a muon neutrino producing a muon track with an energy greater than $E^{\rm thr}_{\mu}$ inside of the volume of IceCube is given by:
\begin{equation}
P_{\nu_{\mu}\rightarrow {\rm track}} \simeq \rho N_A \int^{1-(E^{\rm thr}_{\mu}/E_{\nu})}_{0} R_{\mu}((E_{\nu}(1-y))  \frac{d\sigma}{dy} dy.
\end{equation}

To make a comparison between the rate of shower and muon track events, we consider a spectrum of the form $dN_{\nu}/dE_{\nu} \propto E_{\nu}^{-2}$, and with the flavor ratios predicted from the decays of positively charged mesons, after accounting for oscillations ($\nu_e:\nu_{\mu}:\nu_{\tau}:\bar{\nu}_e:\bar{\nu}_{\mu}:\bar{\nu}_{\tau} \approx 2:1:1:0:1:1$).\footnote{Electron anti-neutrinos will be produced through oscillations of $\bar{\nu}_{\mu}$ and $\bar{\nu}_{\tau}$ for non-zero values of $\theta_{13}$. Thus the rate of showers at the energy of the Glashow resonance (6.3 PeV) could be used as a probe of this oscillation parameter~\cite{Bhattacharya:2011qu}.} Considering only neutrinos with energies in excess of 1 PeV, we find that for every fully contained shower ($V_{\rm eff}\approx 0.1$ km$^3$), one expects 33 (11) muons tracks with more than 100 TeV (1 PeV) of energy to travel through the volume of IceCube. We thus conclude that if these two fully contained PeV showers are neutrino induced, then the existing data set from IceCube should contain tens of PeV muon track events, as well as tens of partially contained PeV showers. Such events would be very valuable for searching for correlations in direction (for track events) and time with known gamma-ray bursts.

We also note that it may be possible for IceCube to identify PeV-scale events which are unique to tau neutrinos~\cite{pakvasalearned}. Following the approach of Ref.~\cite{beacom}, and for the same assumptions as in the previous paragraph, we estimate that for every fully contained PeV shower, one should expect approximately 0.7 lollipop events and 0.25 double bang events.\footnote{Here, we have required only that the showers involved in these events are initiated within the volume of IceCube, and are not necessarily fully contained.} While these are not particularly high rates, even one or two tau-unique events would be very helpful in confirming an astrophysical nature of the events in question.

\bigskip

\section{Summary and Conclusions}
\label{summary}

Experiments such as IceCube were motivated in large part by the connection between the observed cosmic ray spectrum and high-energy cosmic neutrinos. In particular, if a significant fraction of high-energy cosmic rays interact via photo-meson interactions, then it should be possible to observe the resulting diffuse flux of very high-energy cosmic neutrinos. With the observation of two showers with energies of $\simeq$1 PeV at IceCube, we may very well be witnessing the first light in the field of high-energy neutrino astronomy. And if so, the energy and flux implied by these events is strongly suggestive of a connection with the observed cosmic ray spectrum. 

In this paper, we have considered a number of possible sources and mechanisms that could produce PeV neutrinos in a quantity capable of accounting for the two events observed by IceCube. We find that photo-meson interactions of $\sim$10-100 PeV protons to be the most promising possibility, which could plausibly be realized in a variety of astrophysical sources, including gamma-ray bursts (GRBs), active galactic nuclei (AGN), and starburst galaxies. Due to the observed break at 0.1-1 MeV in the spectra of typical GRBs, one expects the resulting neutrino spectrum from such objects to peak at the PeV scale; in fact we find that GRBs can naturally explain the two observed events. More specifically, we find that a neutrino flux comparable to that implied by the two events is predicted for a wide range of assumptions regarding the redshift distribution, luminosity function, and other physical characteristics of GRBs. In contrast, the photon spectra in typical AGN models peak at much lower energies, and thus neutrino spectra are predicted which peak at energies well above those of IceCube's two events. In fact, searches for even higher energy showers in IceCube's existing data set may be capable of excluding some of such models. Cosmic ray collisions in starburst galaxies are also a promising possibility, but it is not clear whether neutrinos with energies as high as 1 PeV can be generated in such sources. 

Lastly, we point out that the two fully contained shower events observed by IceCube should be accompanied by tens of PeV muon tracks and partially contained showers. A more modest number of events unique to tau neutrinos should also be expected to appear. Thus as IceCube's overall data set is further scrutinized, it should become increasingly possible to determine whether these two events are in fact cosmic in nature and, if so, to begin to constrain their origin.

\bigskip

\acknowledgments{We would like to thank Francis Halzen for helpful discussions. 
This work has been supported by the US Department of Energy.}

\end{document}